\documentclass[sigconf]{acmart}

\usepackage{multirow}
\usepackage{tabularx}
\usepackage{float}
\usepackage{url}

\AtBeginDocument{%
  }

\acmConference[CHI '25]{CHI Conference on Human Factors in Computing Systems}{April 26-May 1, 2025}{Yokohama, Japan}

\begin{document}

\title{As Confidence Aligns: Exploring the Effect of AI Confidence on Human Self-confidence in Human-AI Decision Making}

\author{Jingshu Li}
\email{jingshu@u.nus.edu}
\orcid{0009-0006-1576-8487}
\affiliation{%
  \institution{National University of Singapore}
  \country{Singapore}
}

\author{Yitian Yang}
\email{yang.yitian@u.nus.edu}
\orcid{0009-0000-7530-2116}
\affiliation{%
  \institution{National University of Singapore}
  \country{Singapore}
}

\author{Q. Vera Liao}
\email{veraliao@microsoft.com}
\orcid{0000-0003-4543-7196}
\affiliation{%
  \institution{Microsoft Research}
  \city{Montreal}
  \state{Quebec}
  \country{Canada}
}

\author{Junti Zhang}
\email{juntizhang@u.nus.edu}
\orcid{0009-0005-4107-1563}
\affiliation{%
  \institution{National University of Singapore}
  \country{Singapore}
}

\author{Yi-Chieh Lee}
\email{yclee@nus.edu.sg}
\orcid{0000-0002-5484-6066}
\affiliation{%
  \institution{National University of Singapore}
  \country{Singapore}
}


\begin{abstract}
Complementary collaboration between humans and AI is essential for human-AI decision making. One feasible approach to achieving it involves accounting for the calibrated confidence levels of both AI and users. However, this process would likely be made more difficult by the fact that AI confidence may influence users' self-confidence and its calibration. To explore these dynamics, we conducted a randomized behavioral experiment. Our results indicate that in human-AI decision-making, users' self-confidence aligns with AI confidence and such alignment can persist even after AI ceases to be involved. This alignment then affects users' self-confidence calibration. We also found the presence of real-time correctness feedback of decisions reduced the degree of alignment. These findings suggest that users' self-confidence is not independent of AI confidence, which practitioners aiming to achieve better human-AI collaboration need to be aware of. We call for research focusing on the alignment of human cognition and behavior with AI.

\end{abstract}






\maketitle

\newcommand{\rhl}[1]{{#1}}
\section{Introduction}
As artificial intelligence (AI) is increasingly integrated into a range of human decision making processes, research on and support for human-AI decision making is gaining prominence within the human-computer interaction (HCI) and AI communities \cite{ma2023should,wang2023watch,lai2023towards}. 
When collaborating with humans in decision making processes, AI can serve as an advisor, a peer collaborator, or even as a decision-maker \cite{kobis2021bad,trunk2020current,yablonsky2021ai,jain2023effective,lai2022human}.
A key goal of human-AI decision making is for human-AI teams to achieve {\it complementary collaboration}, i.e., joint activity that leads to better outcomes than efforts by either party working alone would achieve \cite{jain2023effective,lai2023towards,ma2023should}. 
One proposed approach to achieving such complementarity is to shift the primary burden of decision making based on team members’ relative levels of uncertainty about their preliminary decisions. For example, a human could delegate the final decision to AI if their own uncertainty is higher; or, an external algorithm could optimize the final decision by weighing each team member’s uncertainty~\cite{li2024overconfident,ma2024you,zhang2020effect,rechkemmer2022confidence,ma2023should}. However, a necessary precondition for such optimization is each team member’s ability to accurately estimate and articulate their decision uncertainty \cite{ma2024you,guo2017calibration}. 

While uncertainty expression of AI can take various forms depending on the model, in this work, we focus on the most common form in human-AI decision making studies and practices---\textit{confidence level}, which is how uncertainty is expressed in classification model (e.g., the model is 80\% confident about the predicted label) \cite{guo2017calibration,zhang2020effect,lai2023towards}. 
This is similar to how humans often express their uncertainty: by assigning a probability of how much they expect their prediction or answer to be correct, i.e. their {\it self-confidence} \cite{yeung2012metacognition,pescetelli2021role,ma2024you}. 

Studies among human decision making groups have observed that the self-confidence of individual members aligns with that of their peers (converging towards a mean value) and then remains aligned in individual decision making after group decision making has ceased \cite{bang2017confidence,Pescetelli2022benefits,fusaroli2012coming}. This phenomenon is known as {\it confidence alignment}.
This points to the potential for AI users' self-confidence to be influenced by AI's expressed uncertainty, and even to align with AI confidence. 
However, previous work theorizing human-AI complementarity by uncertainty \rhl{and studies exploring human self-confidence dynamics in human-AI decision making} have not explored such the possible influence of AI confidence on human self-confidence \cite{li2024overconfident,lai2023towards,zhang2020effect,ma2024you,chong2022human}. 

From the perspective of complementary collaboration, such confidence alignment would be problematic. 
{\it Confidence calibration} involves the correspondence between an individual’s confidence level about their decisions (prior judgment) and those decisions’ accuracy (posterior evidence) \cite{alexander2013calibration,guo2017calibration}. The higher the degree of correspondence, the better the calibration, which is referred to as {\it well-calibrated} or calibrated.
If AI users’ self-confidence indeed aligns with AI confidence, it could lead to miscalibration of the humans’ uncertainty, by changing their self-confidence without changing their decision making capabilities.
Miscalibrated human self-confidence could further undermine appropriate self-reliance and suitable reliance on AI, impairing the outcomes of human-AI decision making \cite{chong2022human,ma2024you}.
And, if user self-confidence aligns closely with AI confidence, it will become unclear which set of decision tasks each excels at, and efforts to optimize delegation between the two will become meaningless. 
Therefore, exploring how AI confidence influences human self-confidence in decision making processes should benefit not only scholarly understanding of the dynamics of human self-confidence within human-AI decision making, but also the finer calibration of human self-confidence and the fostering of effective complementary collaboration in such scenarios.

To this end, we conducted an online randomized behavioral experiment (mixed design, N=270) where participants were asked to perform income prediction tasks \cite{zhang2020effect,misc_census_income_20,hase2020evaluating,chen2023understanding} in collaboration with an AI. 
To facilitate our assessment of whether and how closely the participants’ self-confidence aligned with the AI’s confidence, we designed three decision making task stages as within-subject factors. In the first task stage, we measured participants' baseline self-confidence in 40 individual decision making tasks. Then, in the task second stage, we had participants collaborate with AI in 40 decision making tasks and measured their self-confidence to observe potential alignment. In the third task stage, we asked participants to complete 40 individual decision making tasks again and measured their self-confidence to explore the persistence of confidence alignment.

Meanwhile, previous research has suggested that human-AI decision making involves various collaboration paradigms \cite{lai2023towards,yablonsky2021ai,makarius2020rising}, where AI performs different roles; and in practice, people do not always receive immediate correctness feedback for their decisions \cite{pescetelli2021role}. 
Both collaboration paradigms and the presence or absence of real-time correctness feedback can influence human self-confidence dynamics \cite{chong2022human, perfect2000practice,bang2017confidence}.
\rhl{To thoroughly explore how confidence alignment would be affected under these different situations,} we took the presence of real-time feedback (with and without) and human-AI decision making paradigms (AI as advisor, AI as peer collaborator, AI as decision-maker under human supervision \cite{yablonsky2021ai,lai2023towards,makarius2020rising}) as between-subject factors and had a $2\times3$ experimental design.

Our experimental results indicate that during the human-AI decision making process, participants' self-confidence levels did tend to align with AI confidence levels; and this alignment persisted in individual decision making tasks after the joint decision making process had ended. The presence of real-time feedback, meanwhile, reduced the degree of alignment. Furthermore, the alignment of participant self-confidence with AI confidence influences the calibration of participant self-confidence, which affects the efficacy of human-AI decision making.
In short, during human-AI decision making, human uncertainty is not independent of the uncertainty expressed by AI: a phenomenon that future designers and users of AI should be acutely aware of.

Our work’s key contributions to the HCI community can be summed up as follows:

\begin{itemize}
    \item It shows how AI confidence influences human self-confidence during and after human-AI decision making.
    \item It reveals the changes this alignment brings to the calibration of human self-confidence and the subsequent influences to the outcomes of human-AI decision making.
    \item Its results substantially extend and enrich existing theories of human-AI confidence interactions.
    \item It has important design implications, and includes guidance for future work aimed at improving the quality of human-AI decision making and at using AI to alter levels of human self-confidence.
\end{itemize}

\section{Related Work}

\subsection{Human-AI Decision Making in Current Research and Practices} 
\begin{figure}[t]
\centering
\includegraphics[width=0.5\textwidth]{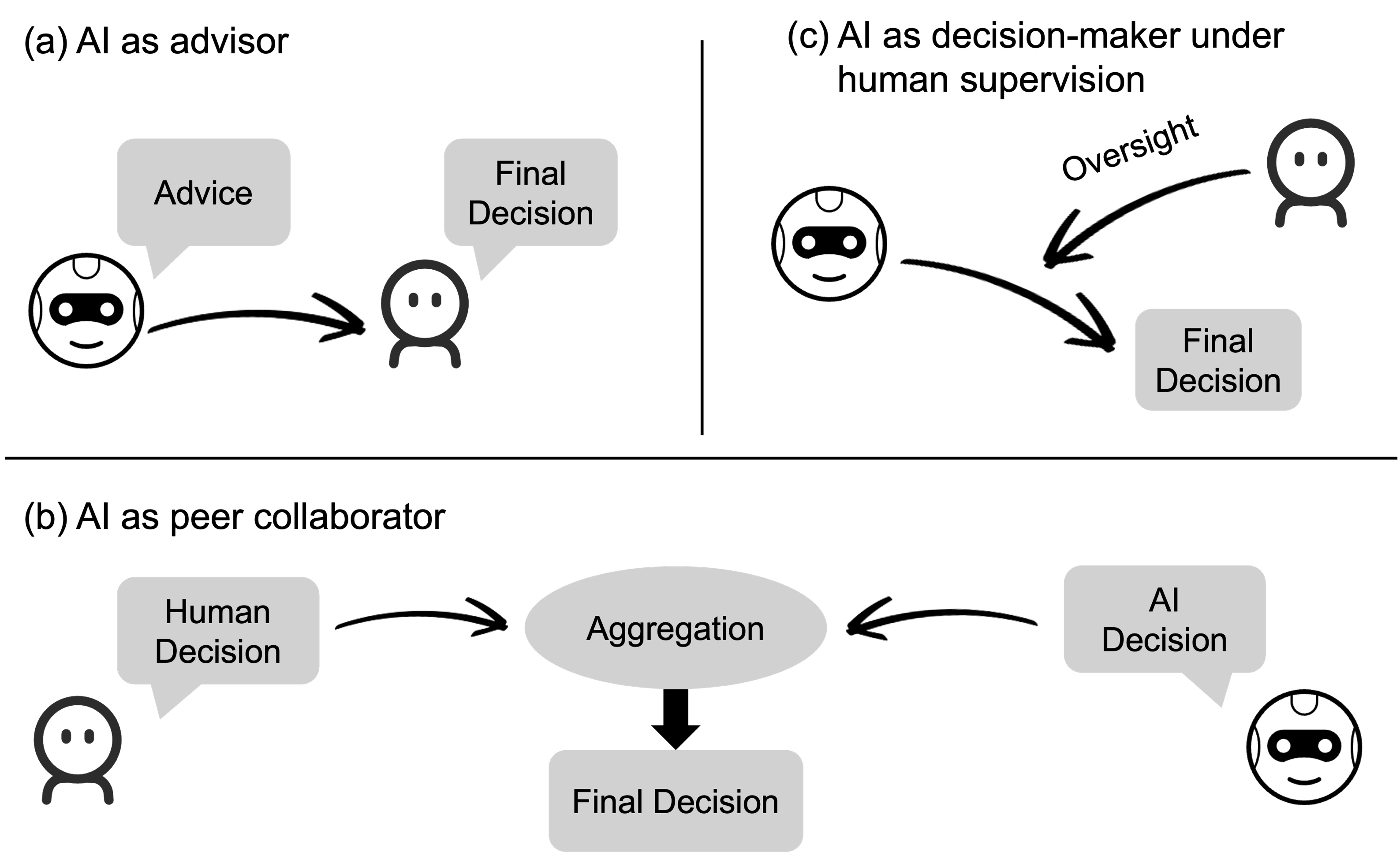}
\caption{Diagrams of three human-AI decision making paradigms.
{\bf (a)}: AI as advisor.
{\bf (b)}: AI as peer collaborator.
{\bf (c)}: AI as decision-maker under human supervision.
}
\label{fig.paradigms}
\end{figure}

Human-AI decision making includes many different paradigms where AI plays various roles within the human-AI group \cite{kobis2021bad,trunk2020current,yablonsky2021ai,jain2023effective,lai2022human}.
This study considers three such paradigms that either have substantial potential for application or are in wide use already (as shown in Fig.~\ref{fig.paradigms}).

Currently, the most widely used such paradigm in HCI and AI research as well as in practical applications involves the AI acting as an advisor. It is generally referred to as {\it AI-assisted decision making} \cite{jain2023effective,li2021algorithmic,bansal2021does,ma2023should,lu2024does,chiang2023two}. In this setup, AI provides humans with information and suggestions, while humans consider AI's recommendations and make the final decisions \cite{pescetelli2021brief,zhang2020effect}. 
For example, in investing, AI provides investors with its investment advice, and investors make decisions by integrating AI suggestions with their understanding of the situation \cite{biran2017human}. 
This paradigm of AI-assisted decision making is anticipated to enhance human capabilities and amplify human intelligence \cite{pescetelli2021brief,leyer2021decision}. The motivation for adopting this paradigm is primarily due to considerations of risk, fairness, ethics, and accountability \cite{zhang2020effect}: As probabilistic models, AI systems cannot guarantee the correctness of specific decisions, and the risks associated with that uncertainty are seen as particularly critical in high-stakes decision domains like finance, healthcare, and law \cite{bansal2021does}.

Alongside the advancement of AI, researchers and designers have begun to explore richer forms of human-AI collaboration \cite{makarius2020rising}, and a mixed-initiative paradigm of human-AI decision making is increasingly being researched  with \cite{puranam2021human,jain2023effective,pescetelli2021brief}. In it, AI is treated as a peer collaborator of equal status to the human decision-maker(s): and their joint decisions are arrived at collectively through various aggregation mechanisms (e.g., choosing the decision with the highest confidence level) \cite{pescetelli2021brief,puranam2021human}. For instance, in stock selections, both humans and AI can provide suggestions for stock prices, and the final decision depends on the degree of agreement between human and AI recommendations \cite{puranam2021human}. 

Lastly, following a comprehensive assessment of risks and benefits, some researchers and practitioners have begun to explore a high-automation paradigm in which decision making is automated by AI, with humans acting as supervisors \cite{markus2017datification,yablonsky2021ai,parry2016rise}. 
That is, AI analyzes data and makes decisions within a framework of human governance and oversight \cite{yablonsky2021ai}. 
While this paradigm can be adopted when AI can reliably make autonomous decisions that are as good as or better than human decisions \cite{pescetelli2021brief,holcomb2018overview}. 
Considering that in real-world applications there may be complex tasks and corner cases that AI cannot address, human observation and supervision, as well as timely intervention in such special circumstances, remain necessary \cite{markus2017datification}.
In most cases, AI will autonomously make decisions, with humans merely observing AI behavior; humans intervene only when anomalies are reported.
This paradigm has several strengths, including higher effectiveness, and eliminating irrelevant sociocultural constraints and cognitive biases \cite{parry2016rise}. 
Some applications of this paradigm already exist, such as Waymo’s autonomous taxis, where AI is responsible for all normal driving decisions, while remote humans intervene to provide assistance in exceptional or complex situations \cite{fairfield2017remote}.


\subsection{AI Confidence, Human Self-confidence, and Complementary Collaboration in Human-AI Decision Making}
The uncertainty estimation of AI can be expressed in various forms, such as confidence levels in classification models and confidence intervals in regression models \cite{ghosh2021uncertainty, lai2023towards}. 
In this paper, we focus on AI confidence levels, the most common form of AI uncertainty estimation in past research on human-AI decision making \cite{lai2023towards, bansal2021does,buccinca2021trust,arshad2015investigating,lee2021human}. 
AI based on classification models can output the conditional probability of a single prediction as its confidence level, reflecting the probability that the prediction is correct \cite{guo2017calibration,zhang2020effect}. 
For example, if the AI reports an 80\% confidence in a predicted label, this implies that it estimates an 0.8 probability that the prediction is correct.
\rhl{As such, AI confidence levels can serve as {\it case-wise} uncertainty indicators, helping humans and external decision making algorithms know when a model is more or less sure, which is different from AI's accuracy that provide {\it overall} performance information over a set of decisions \cite{rechkemmer2022confidence}.}
Humans or external algorithms can then adjust their trust in the AI to appropriate levels and rely on it accordingly based on the level of AI confidence \cite{zhang2020effect,rechkemmer2022confidence}. 
For instance, in some medical and financial decision making practices, AI is required to report its confidence levels alongside its predictions to help humans capture AI uncertainty \cite{rajpurkar2022ai,cao2022ai,weerts2019human,kiani2020impact}. 

Similar to AI confidence level, humans can estimate their uncertainty through their self-confidence levels \cite{yeung2012metacognition}. This estimation of uncertainty can be expressed in natural language, such as "I am not sure," or numerically as "I am 70\% confident" \cite{fusaroli2012coming,bang2017confidence,ma2024you}. Corresponding to the AI confidence level, in this study, participants will express their self-confidence numerically.
Prior research suggests that human self-confidence in decision making is related to metacognition \cite{yeung2012metacognition,kleitman2007self}. Metacognition refers to individuals’ assessment of their own abilities, knowledge, and understanding of task-relevant factors, and their self-confidence can be seen as an outcome of that self-monitoring process \cite{kleitman2007self,keren1991calibration}.
During human-AI decision making process, people’s self-confidence level can be influenced by task-relevant factors such as task difficulty and correctness feedback \cite{Pescetelli2022benefits,chong2022human,perfect2000practice}. 
\rhl{Chong et al.~\cite{chong2022human} found that, in AI-assisted decision making without providing AI confidence, for each task, positive feedback after the task can increase individuals' self-confidence, while negative feedback can diminish it. They also showed that a deterioration in AI accuracy led to reduced human self-confidence, because poorer AI performance increased joint decision errors, then negative feedback lowered human self-confidence~\cite{chong2022human}.}
Human self-confidence level plays an important role in human-AI decision making \cite{chong2022human,ma2024you,bang2017confidence}. On one hand, it governs humans' acceptance or rejection of AI predictions: individuals are more likely to accept AI predictions when their self-confidence is low \cite{chong2022human}. On the other hand, it affects both humans' self-reliance and external algorithms' reliance on humans. 

Complementary collaboration can be realized by accounting for the confidence levels of both humans and AI \cite{li2024overconfident,ma2024you,zhang2020effect,rechkemmer2022confidence,ma2023should}. Under ideal conditions, the optimization of complementary collaboration entails the final decision makers (either humans or algorithms) having appropriate reliance on human and AI decisions according to their confidence levels---using AI predictions for final decisions when AI's confidence level is higher than that of the human, and relying on human judgment when the human's confidence level exceeds that of AI \cite{li2024overconfident,rechkemmer2022confidence}. Achieving such optimization requires calibrated confidence levels of humans and AI \cite{lai2023towards,ghosh2021uncertainty}. 

However, both AI confidence and human self-confidence are facing the miscalibration problem \cite{guo2017calibration,Pescetelli2022benefits}. On the AI side, the confidence calibration of machine learning models is challenging, with many models being overconfident---having confidence levels that exceed their actual accuracy---while others may be underconfident \cite{guo2017calibration,xiong2023can,wang2021confident}. Past research has indicated that these problems are linked to model capability and regularization \cite{guo2017calibration}. 
On the human side, calibrating self-confidence is also challenging because it is influenced not only by the difficulty of the decision tasks themselves, but also by various socio-economic factors such as gender, occupation, psychological health, and so on \cite{broihanne2014overconfidence,huq1988probabilistic,mann1998cross,campbell2004narcissism,Pescetelli2022benefits}. 
Lack of calibration of either party’s confidence level can undermine the efficacy of complementary collaboration between humans and AI. Thus, some recent work has focused on calibrating AI confidence as well as human self-confidence to remedy the miscalibration problems \cite{ma2024you,guo2017calibration}. For instance, Ma et al. \cite{ma2024you} have explored how cognitive interventions, rewards, and real-time feedback can help humans calibrate their self-confidence during human-AI decision making.

\subsection{Confidence Alignment in Decision Making}
Previous research has shown that in decision making processes involving groups of two or more humans, the members’ respective levels of self-confidence tend to come into alignment and approach uniformity \cite{fusaroli2012coming,bang2017confidence,Pescetelli2022benefits}. For instance, in an experiment on dyadic cooperative decision making where dyad members can communicate their decisions and confidence with each other, by encoding the expression of confidence in speech, researchers found that the levels of verbally expressed confidence among dyads tended to converge \cite{fusaroli2012coming}. Most dyads in the experiment also converged to the same set of functional expressions for their confidence \cite{fusaroli2012coming}. Similarly, Bang et al.~\cite{bang2017confidence} also observed that in cooperative psychophysical decision making tasks where confidence was digitally reported, the average confidence levels of group members were more similar when performing tasks cooperatively than when they performed them individually; and this phenomenon occurred irrespective of interpersonal differences in baseline accuracy. 
Recent research has further indicated that in perceptual decision tasks where participants are not required to make cooperative decisions but can see each other's decisions and confidence levels, dyads' digitally expressed confidence also display alignment, and the influence of such alignment can persist in individual decision tasks after their interaction \cite{Pescetelli2022benefits}.

The confidence alignment phenomenon is thought to be a result of humans imitating the confidence of their peers \cite{bang2017confidence}. 
Imitation can be defined as "action that copies the action of another more or less exactly, with or without intent to copy" \cite{english1958comprehensive}, encompassing concepts like behavioral contagion, conformity, social pressure, and social facilitation \cite{wheeler1966toward}. 
Individual behaviors can spread and be imitated from one person to another through observation and interaction \cite{wheeler1966toward,zajonc1987convergence}. 
Furthermore, neurological evidence suggests that people can imitate others' risk preferences by observation \cite{suzuki2016behavioral}. 
Through imitating the behaviors of others, individuals can expand their perceptual and cognitive capabilities at minimal cost \cite{bonabeau1999swarm,eberhart2001swarm}. 
In cooperative decision making contexts specifically, previous studies suggest the confidence alignment is the result of unconscious behavioral imitation in common situations and the intended imitation aimed at adapting to each other's confidence \cite{fusaroli2012coming,bang2017confidence,Pescetelli2022benefits}. 

\section{Research Questions}
Inspired by research on confidence alignment among humans, we propose that in human-AI decision making, human decision-makers' self-confidence may also be influenced by, and align with, AI confidence. 
This idea is supported by a separate body of prior research findings that humans can imitate the behaviors and viewpoints of AI and robots \cite{kobis2021bad, hertz2018under, vollmer2018children, hertz2019mixing,branigan2006alignment}.
The Computers are Social Actors (CASA) theory provides an explanation for these phenomena \cite{nass_CASA, nass_mindlessness}. This theory posits that people naturally and unconsciously use social and interpersonal heuristics to interact with computers \cite{nass_CASA, nass_mindlessness}, and humans' imitation can also be seen as a type of social and interpersonal heuristic, automatically triggered by certain social cues from the computer \cite{hertz2018under, vollmer2018children, hertz2019mixing}.

Previous studies of human-AI decision making have used AI confidence levels to help humans understand AI uncertainties and effectively calibrated AI confidence and human self-confidence to promote complementary collaboration \cite{zhang2020effect,rechkemmer2022confidence,li2024overconfident,lai2023towards,ma2024you, chong2022human}. 
To date, however, our understanding of the influence of AI confidence on human self-confidence remains limited.
Investigating this issue could help researchers and developers better comprehend the dynamics of human self-confidence in human-AI decision making, and thereby facilitate future efforts to optimize complementary collaboration. Therefore, we ask:

\begin{itemize}
    \item {\bf RQ1}: What is the effect of AI confidence on human self-confidence. Do they align, and if so, to what degree?
\end{itemize}

Meanwhile, referring to the results among humans \cite{bang2017confidence}, in different human-AI decision making paradigms \cite{kobis2021bad,trunk2020current,yablonsky2021ai,lai2023towards}, the differences in the modes of collaboration between humans and AI may alter the effect of AI confidence on human self-confidence. 
Importantly, the real-time correctness feedback can influence human self-confidence \cite{chong2022human,ma2024you,perfect2000practice}. In past research about confidence alignment between humans, the real-time feedback are always provided \cite{fusaroli2012coming,bang2017confidence,Pescetelli2022benefits}. However, the fact is that, in practice, not all decision tasks in human-AI decision making practice involve real-time feedback \cite{pescetelli2021role}. Thus, to boost our findings’ generalizability and to thoroughly explore these different situations, we ask the following two sub-questions:

\begin{itemize}
    \item {\bf RQ1.1}: How do different human-AI decision making paradigms influence confidence alignment?
    \item {\bf RQ1.2}: How does the presence of real-time feedback influence confidence alignment?
\end{itemize}

Without changing human capabilities---more specifically, decision making accuracy---any change in human self-confidence will inevitably affect the correspondence between self-confidence and accuracy, thus altering human self-confidence calibration. This can lead to a series of adverse consequences, such as inappropriate reliance on humans or AI, and damage to the efficacy (accuracy) of human-AI decision making \cite{chong2022human,ma2024you}. To date, the potential effects of self-confidence alignment on self-confidence calibration and its consequences remain to be explored. Investigating this issue can help researchers understand and avoid the calibration problems caused by human self-confidence aligning with AI confidence, promoting complementary collaboration in human-AI decision making. Therefore, we propose the second research question.

\begin{itemize}
    \item {\bf RQ2}: How does the alignment of human self-confidence with AI affect human self-confidence calibration? What are the consequences?
\end{itemize}

Additionally, in AI-assisted decision making paradigms, human decision-makers not only provide their initial decisions at the start of each task but also act as the final decision-makers \cite{lai2023towards,zhang2020effect}. In this process, human confidence in the final decisions, i.e., the joint human-AI decision, is of significant referential value. Exploring the influence of AI confidence on human confidence in final decisions and potential alignment can help researchers further understand the influence of AI confidence on human confidence. Therefore, this study proposes the third research question:

\begin{itemize}
    \item {\bf RQ3}: In AI-assisted decision making, how and how much does human confidence in final joint human-AI decisions align with AI confidence?
\end{itemize}

\section{Method}
To investigate the effects of AI confidence on human self-confidence, we designed and conducted an online randomized behavioral experiment. Our study was approved by the Department Ethical Review Committee of School of Computing, National University of Singapore.

\subsection{Participants}
Participants were recruited on the Connect crowdsourcing platform (CloudResearch \footnote{https://www.cloudresearch.com}) who met the following criteria: (1) residing in the USA, (2) aged between 21 to 60 years (as required by the IRB), and (3) able to participate via a personally owned computing device, with gender balance maintained during recruitment. They were not allowed to take part in the experiment more than once. After excluding those who did not complete the task (either voluntarily returned or dropped out for unknown reasons) and those who failed multiple attention checks, a total of 270 unique participants were involved in our study (45 per condition). According to demographic data provided by the Connect platform, among these participants, 51.1\% were female, the average age was 36.6 (SD=9.4), and 62.2\% had at least an bachelor degree. The basic pay for the experiment was \$6, with an expected duration of 30 minutes. To encourage high-quality performance, inspired by previous research \cite{zhang2020effect,ma2024you}, an additional reward of \$2 was offered for each stage where participants achieved an accuracy rate exceeding 90\%, amounting to a maximum bonus of \$6 over three stages.

\subsection{Experiment Task and AI Model}
In this study, {\it income prediction} was employed as the task for human-AI decision making. Participants were required to predict whether an individual's annual income would exceed \$50,000 (roughly the medium income at the time of data collection) based on demographic and employment information. The task utilized data from the Adult Income dataset of the UCI Machine Learning Repository \cite{misc_census_income_20}, which contains 48,842 instances. Each instance is described by 14 attributes, including age, occupation, gender, among other demographic information. The annual income for each instance is binarized, indicating whether it exceeds \$50,000. This task has been used in several previous studies on human-AI decision making \cite{hase2020evaluating,zhang2020effect,ma2023should,ghai2021explainable,chen2023understanding}. It requires low domain-specific expertise, making it suitable for online randomized experiments with participants who have undergone necessary training \cite{ghai2021explainable,ma2023should}. To ensure reasonable task difficulty and cognitive load, following the setup of previous research \cite{ma2024you,zhang2020effect}, we selected 8 out of the 14 attributes to present to participants as decision references (based on the importance of attributes), including age, years of education, work class, occupation, marital status, gender, race, and hours worked per week.

According to previous work \cite{rechkemmer2022confidence}, we trained a machine-learning model based on Random Forest \cite{breiman2001random} using samples from the dataset for the human-AI decision making task. After data preprocessing, two-thirds of the original dataset were randomly split to train our machine learning model, while the remaining one-third comprised the tasks assigned to participants in the experiment. In this binary classification experiment, the conditional probabilities predicted by the Random Forest model were directly used as its confidence level.

\subsection{Procedure and Conditions}
\begin{figure*}[t]
\centering
\includegraphics[width=0.95\textwidth]{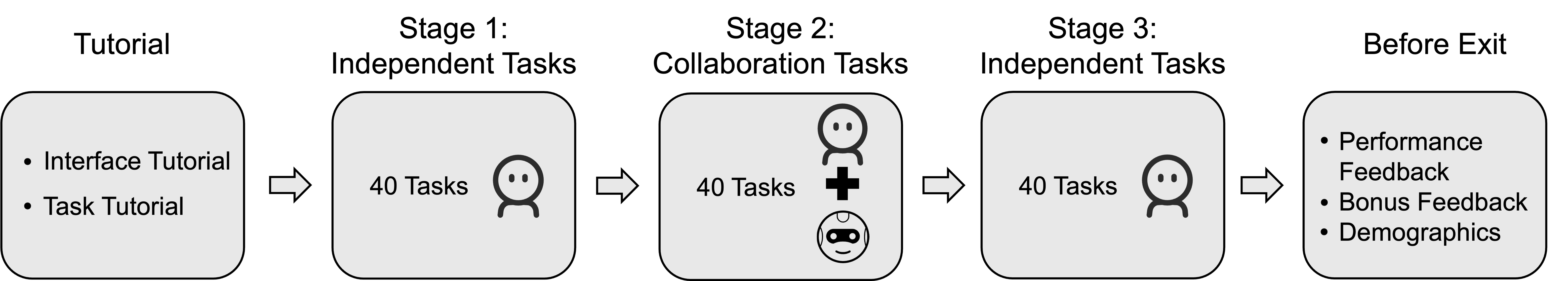}
\caption{Flowchart of the experimental procedure.}
\label{fig.procedure}
\end{figure*}

Our experimental flowchart is presented in Fig.~\ref{fig.procedure}, which was informed by past studies of confidence alignment \cite{fusaroli2012coming,biran2017human,Pescetelli2022benefits}. 
There was a tutorial and three income prediction task stages (comprising 120 different questions in total). 
In the three task stages, the income prediction tasks were the same to each participant and the order of these tasks was randomized within each stage.
In line with the research questions of this paper, the experiment followed a mixed design, with one within-subject factor (3 task stages) and two between-subject factors (presence of real-time feedback and human-AI decision making paradigms).
For the within-subject factor, the first task stage was to measure participants' baseline self-confidence, the second task stage was to measure participants' self-confidence and other relevant variables during human-AI decision making, and the third task stage was to measure participants' self-confidence in individual tasks after human-AI decision making.
For the between-subject factors, there was a 2$\times$3 design: presence of real-time feedback (with real-time feedback, without real-time feedback) $\times$ human-AI decision making paradigms (AI as advisor, AI as peer collaborator, AI as decision maker under human supervision). Below, we describe in detail the procedures and conditions of the experiment.

Specifically, after participants signed the informed consent form, they were directed to a tutorial interface. In tutorial, we explained in writing the objectives of the prediction task and how to use the decision making interface and collaborate with AI. We described each attribute in the profile dataset and demonstrated the binary income distribution corresponding to each attribute. Participants were informed during this stage that achieving an accuracy rate above 90\% in any stage would result in a \$2 reward (cumulative). Before advancing to the next phase, participants were required to correctly complete a fact-check question about the tutorial to ensure they had learned and understood the material.

Participants then entered the first task stage, i.e. independent task stage (stage 1), where they independently completed 40 income prediction questions. Following past research, participants were required to report their prediction and corresponding self-confidence using a slider \cite{Pescetelli2022benefits}. As illustrated, the slider's midpoint served as the boundary: sliding left indicated a prediction of annual income less than \$50,000, with confidence increasing as the slider moved further left (minimum 51\%, maximum 100\%), and vice versa for sliding right. This stage aimed to measure the participants' baseline self-confidence in the income prediction task. Under the conditions with real-time feedback, participants received immediate accuracy feedback after each question (indicating the correctness of the prediction); without real-time feedback, no accuracy feedback was provided.

Subsequently, participants entered the second task stage, i.e. the collaboration task stage (stage 2), where they were required to complete 40 income prediction questions with the AI system. The 40 questions in this stage were manually selected from the test set and different from the questions in stage 1, such that the decision making AI achieved an accuracy of 80\% (32 out of 40 are correct) and an average confidence of 80.40\% (SD=9.64\%). \rhl{The decision making AI could be regarded as close to well-calibrated over these 40 questions.} The specific procedures for this stage, according to the conditions of human-AI decision paradigms in our experiment, are as follows:
\begin{itemize}
    \item \textbf{AI as advisor:} Under this paradigm, our study followed the general paradigms of AI-assisted decision making from past research \cite{zhang2020effect}. For each question, human participants first needed to use a slider to indicate their initial prediction and self-confidence after viewing the question information. Then, the AI's prediction and confidence were displayed to the participants. Finally, participants reported their final decision and confidence in the final decision via the slider, taking into account AI's advice, their initial response, and the question information. 
    Note that in this paradigm, participants reported their confidence level twice for each task. The {\it first} confidence report was regarded as the {\it participants' self-confidence} in their own decision, and the {\it second} confidence report was regarded as their {\it confidence in the joint human-AI decision}.
    In condition with real-time feedback, participants received feedback on the accuracy of their initial decision, AI's decision, and final decision after each question.
    \item \textbf{AI as peer collaborator:} In this paradigm, for each question, participants were first asked to indicate their prediction and self-confidence using a slider after viewing the question information. Then, the AI's prediction and confidence were shown to the participants. Finally, the study adopted the highest confidence rule as the method for aggregating human and AI decisions \cite{bahrami2012failure,koriat2012two}, selecting the decision of the individual with higher confidence (random selection was used for ties) as the final decision, which was then displayed to the participants. In condition with real-time feedback, participants received feedback on the accuracy of their own decision, AI's decision, and the final decision after each question.
    \item \textbf{AI as decision maker under human supervision:} Under this paradigm, for each question, participants acted as a supervisor, observing AI's prediction and confidence. 
    For each task, participants were first required to use a slider to indicate their prediction and self-confidence after viewing the question information (it was {\bf not} a part of decision making under this paradigm, while the purpose was to measure their self-confidence). Subsequently, the AI's prediction and confidence were displayed to the participants. The final decision would automatically use the AI's prediction, and participants would {\bf not} make the final decision.
    Note that this setup reflects the common case in this paradigm, where humans only observe AI making decisions. It does not include cases where humans intervene due to anomalies which do not occur frequently. Because under this paradigm, the decision making process when humans do intervene is essentially the same as when AI acts as an advisor.
    This design was inspired by observation tasks from past research \cite{biran2017human,Pescetelli2022benefits}. 
    In condition with real-time feedback, participants received feedback on the accuracy of their own decision and AI's decision after each question.
   
\end{itemize}

After completing the tasks in stage 2, participants entered the third task stage, i.e. another independent task stage (stage 3). This stage also included 40 income prediction questions (different from stage 1 and stage 2), which participants had to complete without receiving AI assistance. The decision making process and the settings for real-time feedback were the same as in the stage 1. The purpose of this stage was to detect any potential persistent effects of AI confidence on human self-confidence. 

Upon completion of all stages, regardless of the real-time feedback conditions, participants received feedback on their accuracy for each stage, as well as information on whether they received a bonus and the amount of the bonus (for stage 2, when AI acted as an advisor or peer collaborator, the accuracy feedback and bonus determination were based on the accuracy of the final decision of the human-AI team; when AI acted as a decision maker under human supervision, it was based on the accuracy of the human decision). This study also collected participants' demographics.

\subsection{Experimental Interface}
\begin{figure*}[t]
\centering
\includegraphics[width=0.95\textwidth]{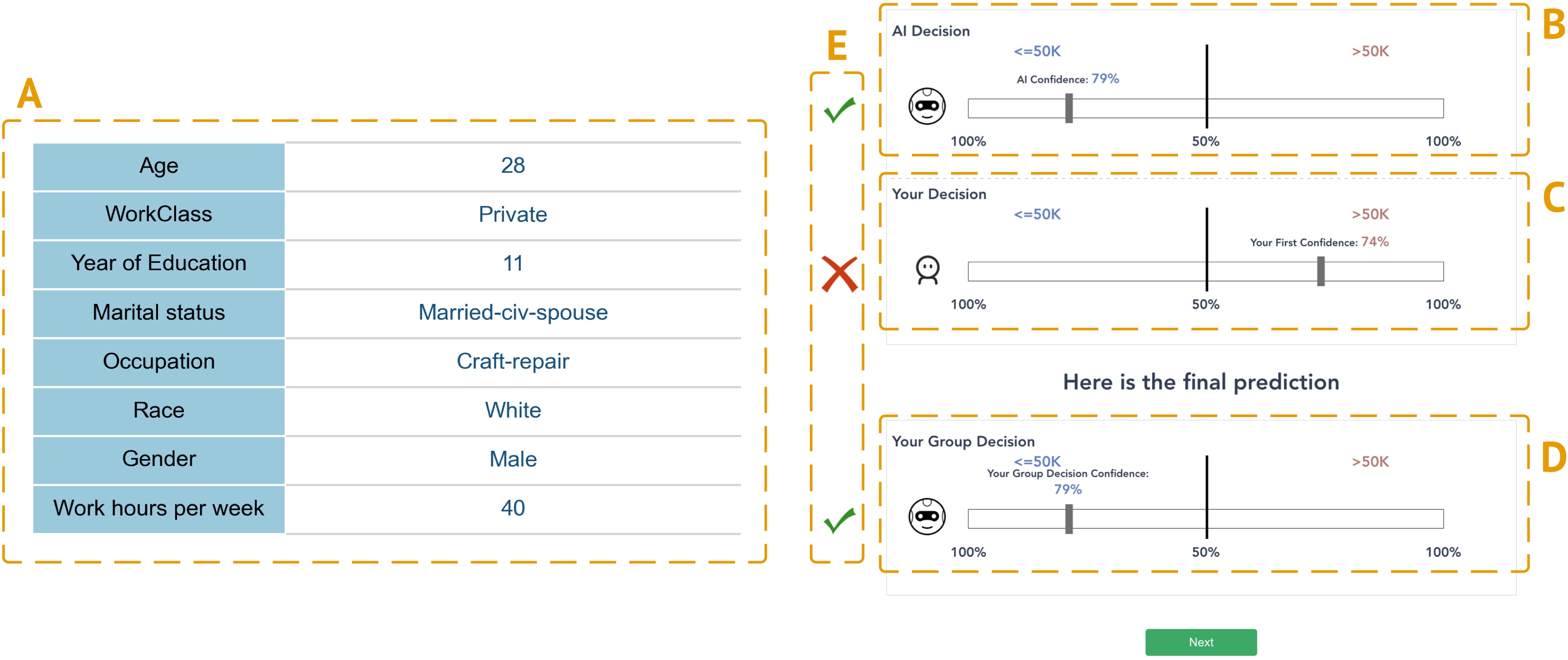}
\caption{Interface of the income prediction task (an example on a task instance from stage 2 and under the paradigm where AI acts as a peer collaborator). 
{\bf A}: The profile table of the income prediction task including 8 attributes. 
{\bf B}: AI prediction and confidence level about the task. They are only presented in stage 2, after users report its prediction and self-confidence at first.
{\bf C}: Users' prediction and self-confidence level about the task. At the beginning of each task from each stage, users need to report and submit their decision and self-confidence here at first. 
{\bf D}: The final decision and confidence about it, only applying to stage 2. For the paradigm where AI acts as an advisor, users need to make the final decision here. For the other two paradigms, the system would make the final decision according to the rules and present the result here.
{\bf E}: Real-time correctness feedback for users' decision (stage 1, 2 and 3), AI's decision (stage 2), and the joint final decision (stage 2): under conditions with real-time feedback, it would be displayed after users submit their own decisions in stages 1 and 3, and after the final decision is made in stage 2.
}
\label{fig.interface}
\end{figure*}

As shown in Fig.~\ref{fig.interface}, the online experimental interface of this study was implemented using the \textit{JavaScript} framework \textit{Vue.js}. The system captured participant decisions and self-confidence levels, and utilized \textit{MySQL} for backend storage management. Additionally, this study employed the Qualtrics~\footnote{\url{https://www.qualtrics.com/}} online survey platform for obtaining informed consent and delivering tutorials.

\subsection{Measurement}

\paragraph{Participants' Self-confidence} The self-confidence of participants at each stage was represented by the average across the self-confidence levels they report for their own decisions within that stage.

\paragraph{Confidence Alignment} Following previous experiments, for each stage, we utilized the absolute difference in mean confidence between AI ($C_{AI}=80.40\%$) and participants $C_{Human}^{Stage}$ in that stage, expressed as $|C_{Human}^{Stage}-C_{AI}|$ to measure the degree of confidence alignment \cite{bang2017confidence,Pescetelli2022benefits}. For short, it was called {\it absolute confidence difference}. A smaller value indicates a higher degree of confidence alignment. 

\paragraph{Self-confidence Calibration} As shown in Equation~\ref{ECE}, following previous research \cite{guo2017calibration,ma2024you}, we used the expected calibration error (ECE) as a measure of participants' self-confidence calibration. Initially, we divided the domain of self-confidence into $M$ bins of equal width (in this paper, $M=4$). The confidence levels reported by participants for $N$ predictions were divided into these $M$ bins. For each bin $B_m$, the average confidence $conf(B_m)$ and accuracy $acc(B_m)$ were calculated, and their difference was taken as the absolute value. Finally, the absolute differences in each bin were averaged, weighted by the number of predictions in corresponding bin $|B_m|$, to calculate the ECE. A smaller ECE value indicates better calibration of self-confidence.

\begin{equation}
    ECE=\sum_{m=1}^{M}\frac{|B_m|}{N}|acc(B_m)-conf(B_m)|
    \label{ECE}
\end{equation}

\begin{figure*}[t]
\centering
\includegraphics[width=0.95\textwidth]{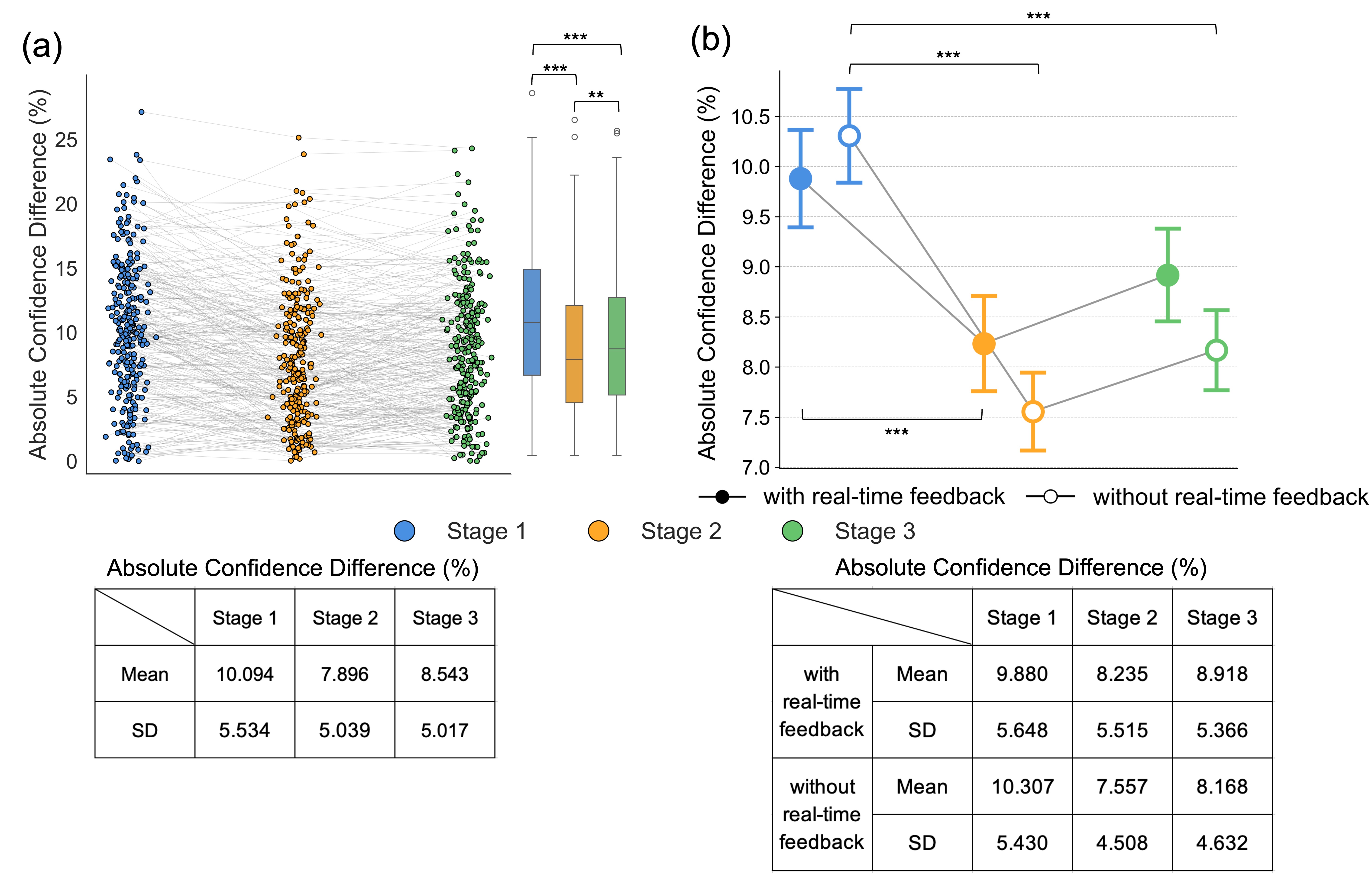}
\caption{Absolute confidence difference between participants' self-confidence and AI confidence (80.40\%). The smaller the difference, the higher the alignment of participants' self-confidence with AI confidence. \rhl{The mean values and standard deviations of the absolute confidence difference in each stage are shown in the tables below corresponding subfigures. The significance levels are labeled ($p<0.05$: *, $p<0.01$: **, $p<0.001$: ***).}
{\bf (a)}: The jitter plot on the left displays the absolute confidence difference for each participant from stage 1 to stage 3, with results from the same participants connected by lines. The box plot on the right illustrates the distribution of the absolute confidence differences at each stage, where the center line represents the median, the box boundaries are the upper and lower quartiles, and the whiskers extend to the extreme points, with outliers also marked.
{\bf (b)}: Line plot about the absolute confidence difference across three stages and two different real-time conditions. The points represent mean values, and the error bars represent one standard error.}
\label{fig.align}
\end{figure*}

\paragraph{Inappropriate Reliance and Human-AI Decision Making Efficacy}
To explore the effects of the change of participant self-confidence calibration, we measured the inappropriate reliance behaviors on AI when it served as an advisor, the inappropriate reliance of decision making mechanism on participants and AI when AI acted as peer collaborator, and the efficacy of human-AI decision making across both paradigms. As the decision making is entirely conducted by AI under human supervision, with a constant accuracy of 80\%, we do not consider the consequences of changes in human self-confidence calibration in this paradigm.

Inspired by previous research \cite{ma2024you,li2024overconfident}, the measurement of inappropriate reliance behaviors on AI when it served as advisor includes the {\it over-reliance}, which is the percentage of tasks where the participants' first decision is correct, AI's prediction is incorrect, yet the final decision is incorrect; and the {\it under-reliance}, which is the percentage of tasks where the participants' first decision is incorrect, AI's prediction is correct, yet the final decision is incorrect.

When AI acts as a peer collaborator, the measurement of inappropriate reliance of decision making mechanism includes {\it over-reliance on AI}, which is the percentage of tasks where the participants' decision is correct, AI's prediction is incorrect, yet the final decision is incorrect; and {\it over-reliance on human}, which is the percentage of tasks where the participants' decision is incorrect, AI's prediction is correct, yet the final decision is incorrect.

For the above paradigms, the measurement of human-AI decision making efficacy is defined as the {\it accuracy} of the joint final decision in stage 2.

\paragraph{Participants' Confidence in Joint Human-AI Decision}
To address {\bf RQ3}, which explores the alignment between participant confidence in the final decision, i.e., the joint human-AI decision, and AI confidence when AI serves as an advisor, we calculated the mean confidence level reported by participants in the final decision during stage 2 in this paradigm. Then we also utilized the absolute difference in mean confidence between AI and participants' confidence in the final decision to measure the degree of confidence alignment \cite{bang2017confidence,Pescetelli2022benefits}.





\section{Results}
In this section, we sequentially report our findings to address {\bf RQ1}-{\bf RQ3}. The participants achieved an average accuracy of 63.85\% ($SD=5.32\%$) across 120 tasks without AI's assistance.

\begin{figure*}[t]
\centering
\includegraphics[width=0.76\textwidth]{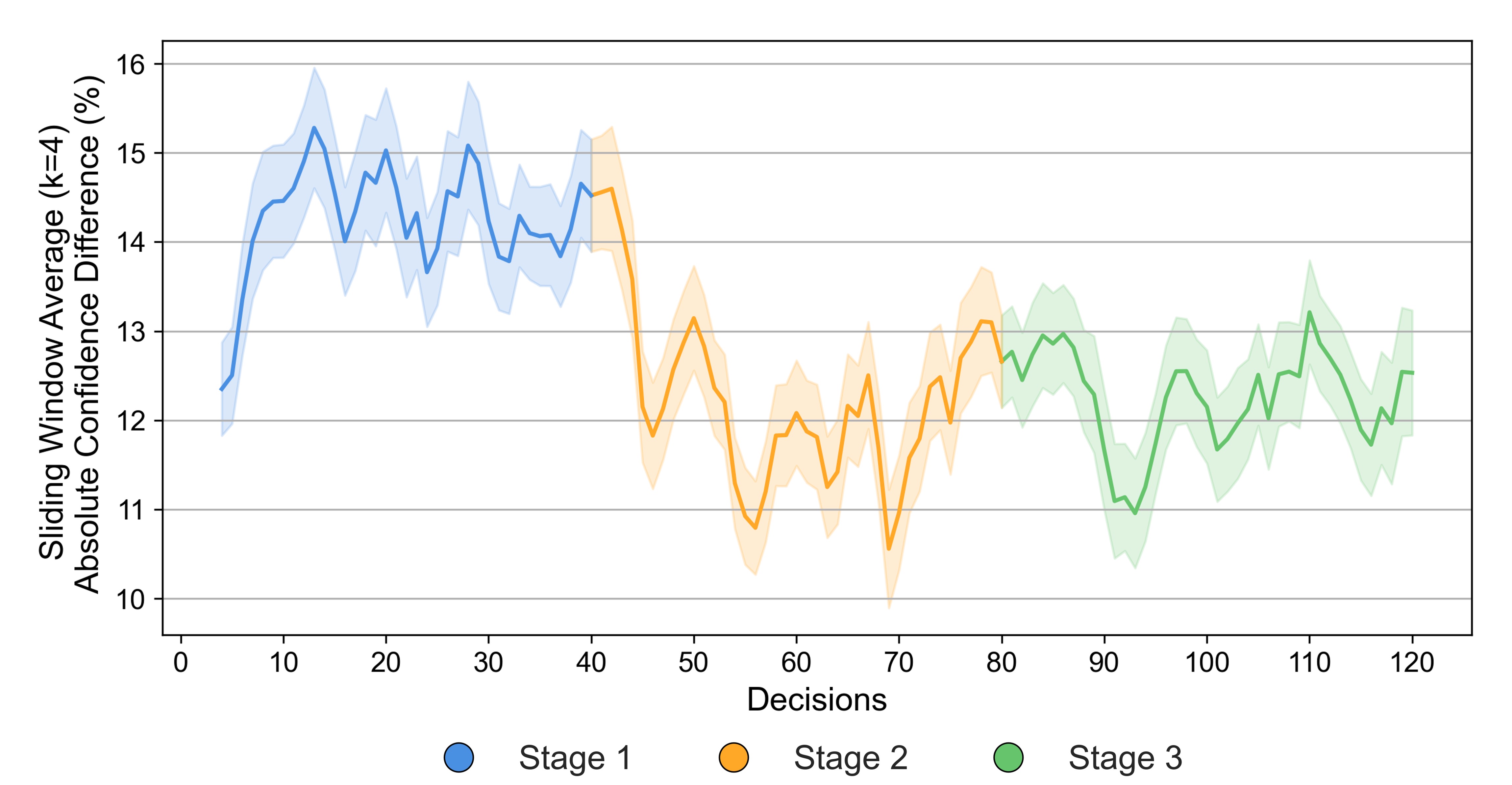}
\caption{
\rhl{Sliding window average (k=4) of the decision-by-decision absolute confidence difference between participants' self-confidence and AI's mean confidence over 120 decision tasks in this experiment with 95\% confidence interval displayed. The absolute confidence difference reduced at stage 2, where participants were given AI confidence levels.}}
\label{fig.sliding}
\end{figure*}

\subsection{Participants' Self-confidence Aligned with AI Confidence (RQ1)}
Repeated measures ANOVA was employed to see the effect of AI confidence on participants' self-confidence and the effects of human-AI decision making paradigms and presence of real-time feedback on this process. 
In the repeated measures ANOVA, three distinct task stages were treated as a repeated measures factor (within-subject factor), with human-AI decision making paradigms and the presence of real-time feedback as between-subject factors. The absolute difference between participants' self-confidence and AI confidence served as the dependent variable.

As the sphericity test for the repeated measures ANOVA was violated ($W=0.926$, $\chi^2=20.189$, $p<0.001$), a Huynh-Feldt correction was applied. Results indicated that the main within-subject effect of the repeated measures factor on the mean difference between participants' self-confidence and AI confidence was significant ($F(1.875, 495.024) = 45.181$, $p<0.001$, $\eta^2=0.031$, Partial $\eta^2=0.146$), as shown in Fig.~\ref{fig.align} (a). Holm-Bonferroni method post-hoc analysis revealed that the absolute confidence difference at stage 2 was significantly lower ($t=-9.142$, $p<0.001$, Cohen's $d=-0.421$) than at stage 1. The absolute confidence difference at stage 3 was significantly lower ($t=-5.901$, $p<0.001$, $d=-0.297$) than at stage 1 but was significantly higher ($t=3.138$, $p=0.002$, $d=0.124$) than at stage 2. 
\rhl{The mean absolute confidence difference values and standard deviations of each stage are shown in the table under Fig.~\ref{fig.align} (a).}
\rhl{Additionally, Fig.~\ref{fig.sliding} further illustrates the changes in the absolute confidence difference throughout the experimental process: the absolute confidence difference decreased at Stage 2.}

\rhl{These results suggest that {\bf participants' self-confidence levels tend to align with AI confidence levels during the human-AI decision making process}. 
Furthermore, {\bf this alignment, although weakened, still persists in the individual decision making tasks after the conclusion of human-AI decision making}.} 

Furthermore, the interaction effect of the repeated measures factor and the presence of real-time feedback on the absolute confidence difference was significant ($F(1.875, 495.024) = 3.855$, $p=0.024$, $\eta^2=0.003$, Partial $\eta^2=0.014$), as shown in Fig.~\ref{fig.align} (b). 
Holm-Bonferroni method post-hoc analysis showed that, in the absence of real-time feedback, the absolute confidence difference at stage 2 was significantly lower ($t=-8.089$, $p<0.001$, $d=-0.527$) than at stage 1. 
The absolute confidence difference at stage 3 was significantly lower ($t=-5.756$, $p<0.001$, $d=-0.410$) than at stage 1. 
No significant differences ($t=-2.096$, $p=0.195$, $d=-0.117$) were found between stages 2 and 3 in the absence of real-time feedback. 
With real-time feedback, the absolute confidence difference at stage 2 was significantly lower ($t=-4.839$, $p<0.001$, $d=-0.315$) than at stage 1. No significant differences in the absolute confidence difference were found between stage 1 and stage 3 ($t=2.588$,$p=0.092$, $d=0.184$), or between stage 2 and stage 3 ($t=-2.343$, $p=0.159$, $d=-0.131$). 
\rhl{The mean absolute confidence difference values and standard deviations of each stage are shown in the table under Fig.~\ref{fig.align} (b).}
\rhl{For RQ1.2, {\bf the presence of real-time feedback diminishes the alignment of participant self-confidence with AI confidence, both during human-AI decision making and in individual tasks after human-AI decision making.}}

\rhl{The main between-subject effects of human-AI decision making paradigms ($F(2,264)=0.514$, $p=0.599$, $\eta^2=0.003$) and the presence of real-time feedback ($F(1,264)=0.339$, $p=0.561$, $\eta^2=0.001$) were not significant. No other significant interaction effects were observed.
Therefore, for {\bf RQ1.1}, {\bf no significant main effect or interaction effect of human-AI decision making paradigms were observed.}}

\begin{figure}[t]
\centering
\includegraphics[width=0.49\textwidth]{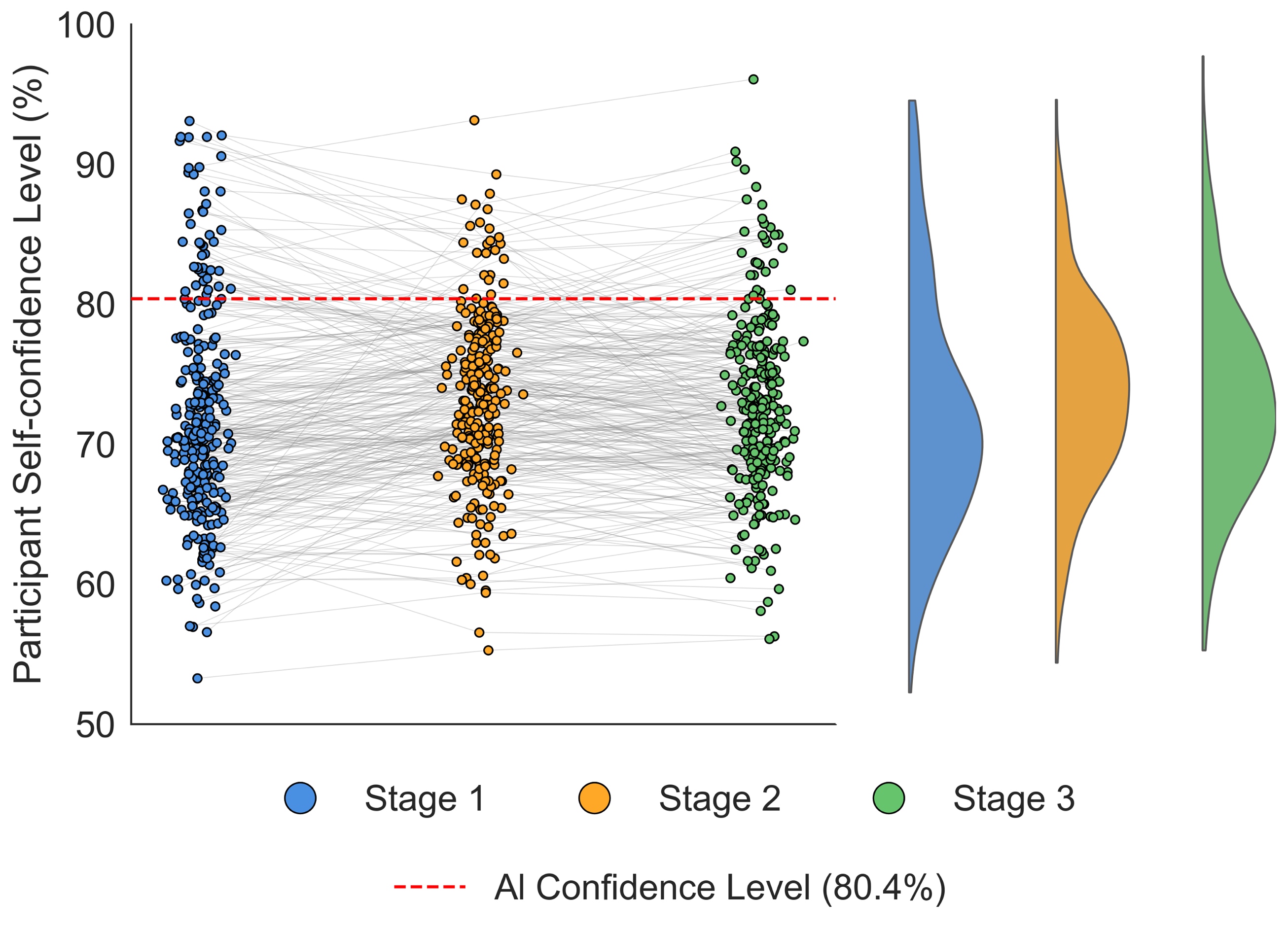}
\caption{Participants' self-confidence level. The jitter plot on the left displays the self-confidence of each participant from stage 1 to stage 3, with results from the same participants connected by lines. The violin plot on the right illustrates the distribution of the absolute confidence differences at each stage. The red dashed line indicates the average AI confidence level in this experiment.}
\label{fig.conf}
\end{figure}

\subsubsection{The Exclusion of Irrelevant Causes}
\label{section:exclusion}
Further analysis was conducted using the same factors, with participants' accuracy as the dependent variable in a Repeated Measures ANOVA, which revealed no significant within-subject effects ($F(2,528)=0.135$, $p=0.874$, $\eta^2=2.375\times 10^{-4}$). 
These findings indicate that \rhl{\bf participants' accuracy did not significantly vary across task stages, preliminarily ruling out the influence of a learning effect}. 

Linear regression analysis showed that no significant linear correlations were found between participants' accuracy and the absolute confidence difference at stage 2 ($r=-0.071$, $p=0.242$) or stage 3 ($r=-0.058$, $p=0.339$). The non-significant linear correlation between accuracy and absolute confidence difference at stages 2 and 3 suggests that \rhl{\bf the alignment of participants' self-confidence with AI confidence during and after the human-AI decision making process is not influenced by the level of participants' own accuracy}.

Additionally, the results shown in Fig.~\ref{fig.conf} also indicate that the alignment of participants' self-confidence with AI confidence \rhl{\bf is not solely driven by a unidirectional change in participants' self-confidence}. During the alignment process from stage 1 to stages 2 and 3, the self-confidence levels of some participants who initially had higher self-confidence than AI confidence decreased, while those of participants who had lower self-confidence than AI confidence increased. This rules out the possibility that the self-confidence alignment is only due to reasons such as AI enhancing people's self-confidence.

\subsection{The Alignment Changed Participants' Self-confidence Calibration and Affected Human-AI Decision Making Efficacy (RQ2)}
\begin{figure}[t]
\centering
\includegraphics[width=0.445\textwidth]{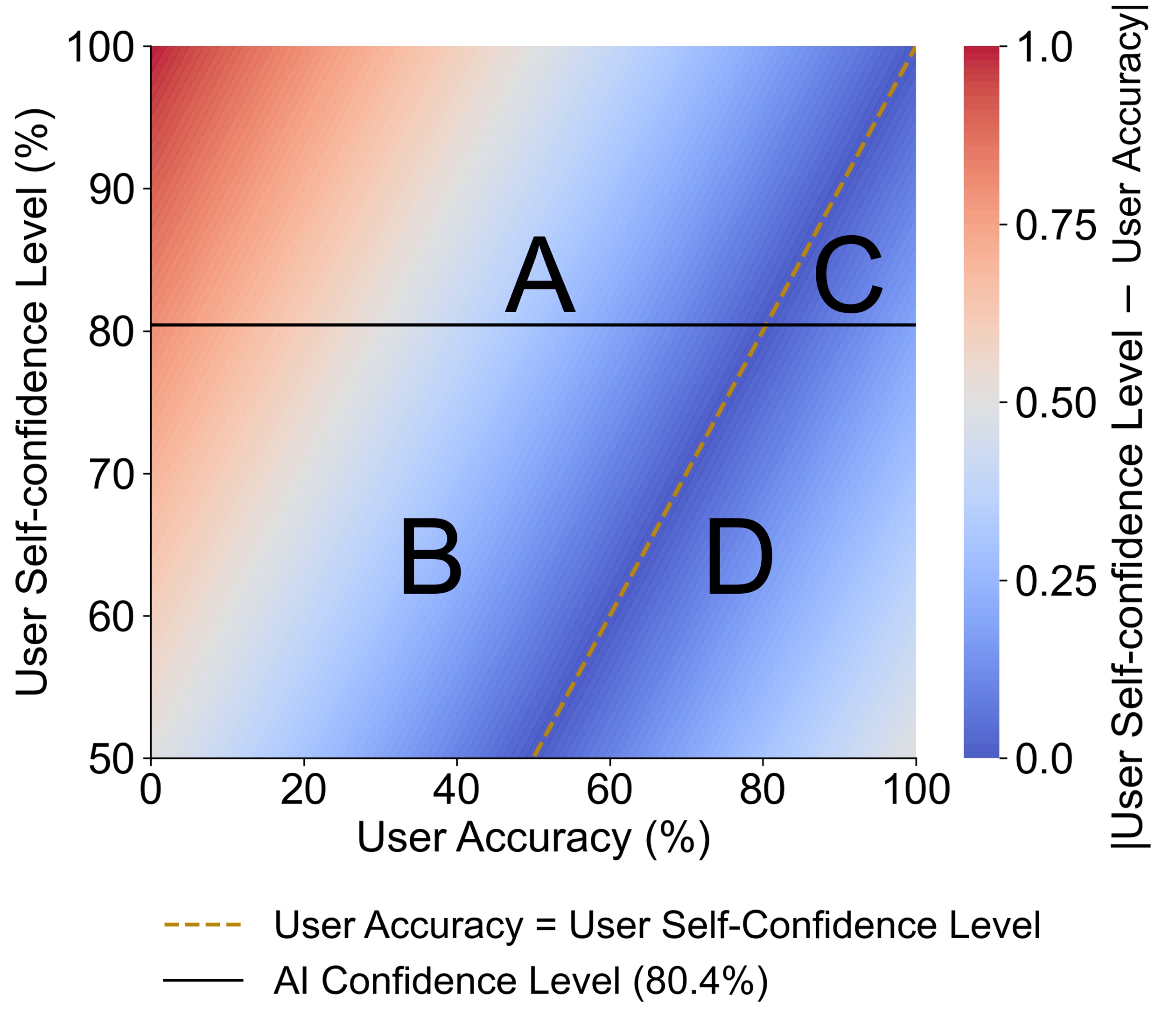}
\caption{A demonstration model used to analyze the influence of the alignment on user self-confidence calibration. The x-axis represents the user's accuracy, while the y-axis shows the user's self-confidence level. The color represents the absolute difference between user's accuracy and self-confidence, serving as a simple method to depict their correspondence, i.e., calibration. The color gradient ranges from blue to red, indicating progressively poorer calibration. The black solid line represents the average AI confidence level in this experiment, and the yellow dashed line indicates the position where user accuracy equals self-confidence, signifying optimal self-confidence calibration in this demonstration model. Users are divided into four regions, labeled A, B, C, and D. \rhl{Type A participants are overconfident and more confident than AI. Type B participants are overconfident but less confident than AI. Type C participants are underconfident but more confident than AI. Type D participants are underconfident and less confident than AI.}}
\label{fig.ece}
\end{figure}

Theoretically, when participants' accuracy remained constant while their self-confidence changed, their self-confidence calibration could also change. As shown in Fig.~\ref{fig.ece}, for participants \rhl{who are overconfident but less confident than AI} ({\it type B}), and those \rhl{who are underconfident but more confident than AI} ({\it type C}), a higher degree of alignment of their self-confidence with AI confidence can result in a greater mismatch \rhl{between their self-confidence and their accuracy}, thereby worsening self-confidence calibration. 
For participants \rhl {who are overconfident and more confident than AI} ({\it type A}), aligning their self-confidence with AI confidence can increase the degree of correspondence \rhl{between their self-confidence and their accuracy}, thus improving their self-confidence calibration. 
For participants \rhl{who are underconfident and less confident than AI} ({\it type D}), \rhl{aligning their self-confidence with AI confidence initially improves self-confidence calibration. As the degree of alignment increases, the improvement diminishes and eventually transitions into a worsening effect.}

\begin{figure*}[t]
\centering
\includegraphics[width=0.85\textwidth]{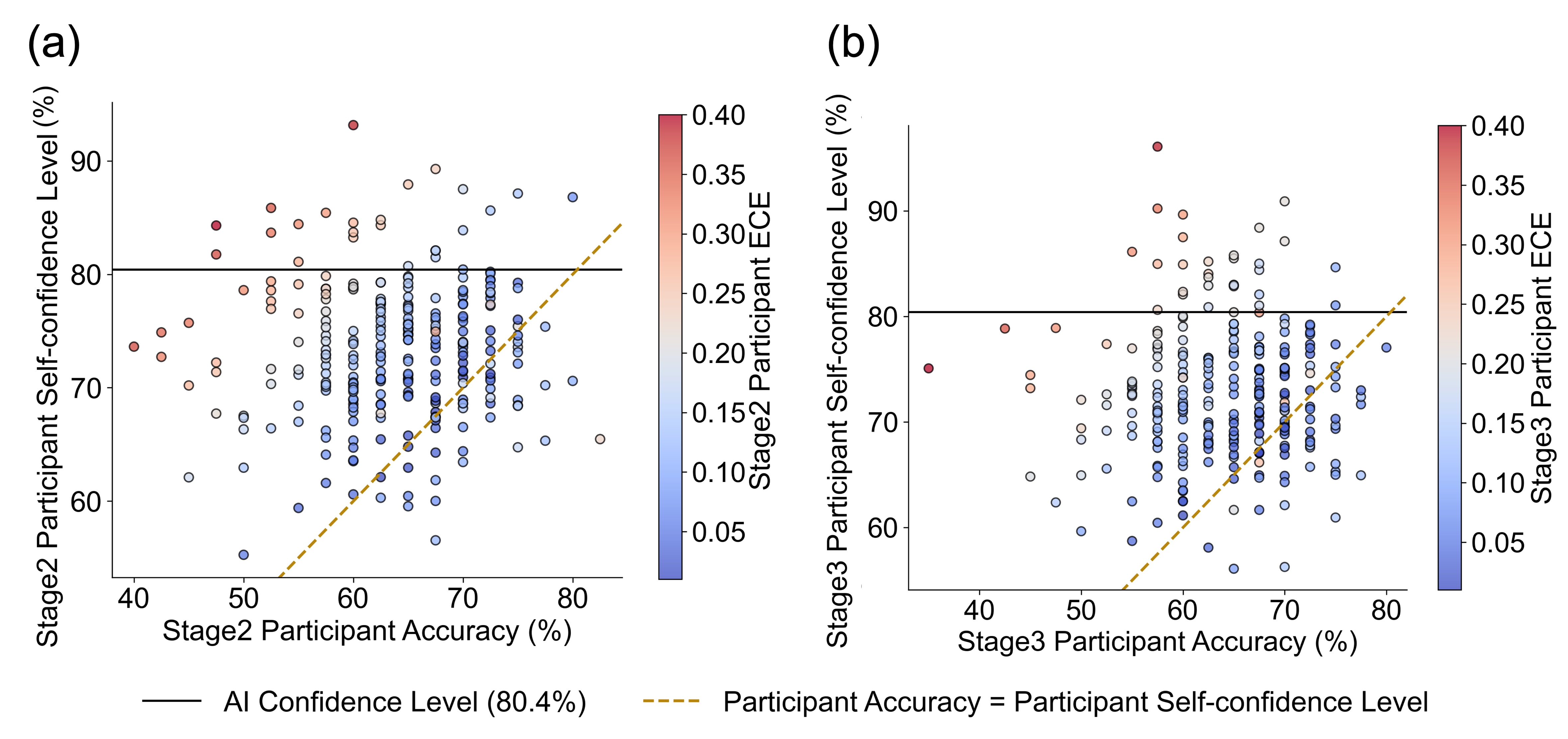}
\caption{Participants' ECE, self-confidence level and accuracy in this experiment. Each point represents a participant. {\bf (a)}: Stage 2's results. {\bf (b)}: Stage 3's results.}
\label{fig.ece2}
\end{figure*}

In this experiment (as shown in Fig.~\ref{fig.ece2} (a)), at {\bf stage 2}, there were 25 {\it type A} participants, 201 {\it type B} participants, 0 {\it type C} participants, and 44 {\it type D} participants. Partial correlation results, controlling for participants' accuracy, indicated a significant positive linear relationship between ECE and absolute confidence difference for {\it type A} participants ($r=0.985$, $p<0.001$), a significant negative linear relationship for {\it type B} participants ($r=-0.797$, $p<0.001$), and a significant positive linear relationship for {\it type D} participants ($r=0.313$, $p=0.041$).

At {\bf stage 3}, there were 29 {\it type A} participants (as shown in Fig.~\ref{fig.ece2} (b)), 194 {\it type B} participants, 0 {\it type C} participants, and 47 {\it type D} participants. Partial correlation results, again controlling for accuracy, revealed a significant positive linear relationship between ECE and absolute confidence difference for {\it type A} participants ($r=0.952$, $p<0.001$), a significant negative linear relationship for {\it type B} participants ($r=-0.722$, $p<0.001$), and a significant positive linear relationship for {\it type D} participants ($r=0.379$, $p=0.009$).

\rhl{The experimental results and analyses are consistent, indicating that {\bf for participants who were overconfident but less confident than AI, aligning their self-confidence with AI confidence degrades their self-confidence calibration. For participants who were underconfident and less confident than AI or overconfident and more confident than AI, the alignment with AI confidence could improve their self-confidence calibration. }}

Linear regression results showed a significant positive correlation between participants' accuracy and self-confidence in {\bf stage 1} ($r=0.172$,$p=0.005$). However, no significant linear correlation were found between participants' accuracy and self-confidence in {\bf stage 2} ($r=0.058$,$p=0.342$) and {\bf stage 3} ($r=0.025$,$p=0.688$).
Comparing to stage 1, the correlation between participants' self-confidence and accuracy is disrupted in stages 2 and 3, indicating poorer overall self-confidence calibration for participants.

Furthermore, when AI served as advisor, in {\bf stage 2}, linear correlation analysis showed that participants' ECE is positively correlated with their under-reliance percentage on AI ($r=0.335$, $p=0.001$), and negatively correlated with their over-reliance percentage on AI ($r=-0.234$, $p=0.027$), as well as negatively correlated with the final accuracy of human-AI decision making ($r=-0.215$, $p=0.042$). 
When AI acted as peer collaborator, in {\bf stage 2}, the results indicated that participants' ECE is positively correlated with the over-reliance percentage of decision making mechanism on AI ($r=0.333$, $p=0.001$), positively correlated with the over-reliance percentage of decision making mechanism on the participant ($r=0.573$, $p<0.001$), and negatively correlated with the final accuracy of human-AI decision making ($r=-0.425$, $p<0.001$).
These results indicate that \rhl{\bf poor participants' self-confidence calibration impairs participants' appropriate reliance on AI and the decision making mechanism's appropriate reliance on both participants and AI, and it can also diminish the efficacy of human-AI decision making}.

\subsection{Participants' Confidence in Joint Human-AI Decision Also Aligned with AI Confidence (RQ3)}
When AI served as an advisor, the results of a repeated measures ANOVA ($F(1, 88) = 49.234$, $p < 0.001$, $\eta^2 = 0.229$), with the presence of real-time feedback as a between-subject factor, revealed that the absolute difference between participant confidence in the final decision and AI confidence ($M=2.975$, $SD=3.838$) was significantly lower than the absolute difference between participant self-confidence in their first decisions and AI confidence ($M=7.936$, $SD=5.211$). The main between-subject effect ($F(1,88)=1.017$, $p=0.316$, $\eta^2=0.004$) of the presence of real-time feedback and its interaction effect ($F(1,88)=3.150$, $p=0.079$, $\eta^2=0.015$) were not significant. 
These results suggest that \rhl{\bf under the AI-assisted decision making, the alignment of participants' confidence in joint final decisions with AI confidence is higher than the alignment of their self-confidence in first decision with AI confidence}. 

Furthermore, another repeated measures ANOVA ($F(1, 88) = 23.860$, $p < 0.001$, $\eta^2 = 0.088$), also with the presence of real-time feedback as a between-subject factor, indicated that when participants’ final decisions agreed with AI predictions, the absolute difference between the mean confidence in the final decision and AI confidence ($M=3.282$, $SD=3.825$) was significantly lower than when the final decisions did not agree with AI predictions ($M=6.823$, $SD=7.154$). 
\rhl{Among all tasks, in 69.056\% of cases, participants' first decision and final decision both agreed with the AI; In 19.056\% of cases, the first decision disagreed with the AI, but the final decision agreed with the AI; In 11.888\% of cases, the final decision did not agree with the AI's predictions.}
Again, the main between-subject effect ($F(1,88)=3.242$, $p=0.075$, $\eta^2=0.022$) of the presence of real-time feedback and its interaction effect ($F(1,88)=0.039$, $p=0.843$, $\eta^2=1.462\times10^{-4}$) were not significant.
\rhl{These suggest that {\bf when participants' final decisions agreed with AI predictions, the alignment of confidence in the final joint decisions with AI confidence is further enhanced compared to when their final decisions disagreed with AI predictions.}}

\section{Discussion}
Our study investigates the phenomenon of human self-confidence aligning with AI confidence and the enduring influence of this phenomenon on human self-confidence. 
The results suggest that in human-AI decision making, human self-confidence levels tend to align with the AI's confidence levels. This alignment persists in individual decision making tasks even after the human-AI collaboration has concluded. 
The presence of real-time feedback reduces the degree of alignment between human self-confidence and AI confidence. This alignment influences the calibration of human self-confidence, which in turn affects appropriate reliance on AI and the overall efficacy of human-AI decision-making processes.
In this section, we summarize these findings and discuss their possible reasons. 

\subsection{On the Results of Human Self-confidence Aligning with AI Confidence}
Our findings reveal a cognitive process in which individuals' self-confidence for a given task is influenced by, and aligns with, the previously observed AI confidence levels during collaborating with AI. 
This process is resembles prior research findings on mutual confidence alignment between humans \cite{fusaroli2012coming, bang2017confidence, Pescetelli2022benefits}, but the alignment of human self-confidence towards AI confidence is unidirectional.
Our results highlight the possibility that human self-confidence—and potential biases within it—are influenced not only by individual thinking styles and socio-economic factors~\cite{bang2017confidence,harvey1997confidence} but also to confidence-related information encountered in human-AI interactions. 
Drawing from Bang et al. \cite{bang2017confidence}, our study suggests that human active or passive imitation to AI is a key reason for the alignment of human self-confidence to AI confidence.
Our analysis did not support the idea that changes in accuracy due to learning effects or other reasons lead to the alignment of human self-confidence, nor did it support the possibility that the alignment is merely due to the presence of an AI collaborator causing an increase in human self-confidence levels.

We also found that the presence of real-time feedback can weaken the alignment of human self-confidence towards AI confidence. 
This is because real-time feedback enables humans to adjust their self-confidence based on their observed correctness~\cite{chong2022human,ma2024you,perfect2000practice}, making their self-confidence more closely with their actual accuracy.
While in our experiment, there was a discrepancy between most participants' accuracy and AI confidence levels. Consequently, the process facilitated by real-time feedback, which aligns human self-confidence with their own accuracy, diminishes the alignment of human self-confidence with AI confidence.

We did not observe a significant effect of different human-AI decision-making paradigms on alignment or its persistence.
Even if participants merely observed AI decisions and confidence, their self-confidence was affected, aligning with findings from previous confidence alignment research between humans \cite{bang2017confidence,Pescetelli2022benefits}. This finding indirectly supports imitation from observation as a cause for alignment.

\subsection{The Influence of Alignment on Human Self-confidence Calibration}
\rhl{The alignment of participants' self-confidence with AI confidence led to a change in their self-confidence calibration. This effect occurred because confidence alignment altered participants' confidence without changing their accuracy. 
In our experiment, some of individuals experienced a deterioration in self-confidence calibration, while the others showed improvement in self-confidence calibration.}
For the deteriorated ones, the alignment causes their self-confidence to move closer to AI confidence but away from their actual accuracy, thus reducing the degree of correspondence between self-confidence and accuracy, and naturally worsening their calibration. 
Conversely, for the improved ones, the alignment causes their self-confidence to move closer to both AI confidence and their accuracy, thus improving their calibration. Overall, the influence of alignment on human self-confidence calibration depends on the numerical relationships among AI confidence, human self-confidence, and human accuracy. 

In this study, the majority belonged to the group for whom self-confidence alignment worsened self-confidence calibration. \rhl{This was because most participants were {\it type B}, which were overconfident but their self-confidence was lower than AI's (Fig.~\ref{fig.ece}). Consequently, in stages 2 and 3, no significant correlation was found between participants' accuracy and self-confidence. 
We had few {\it type A} participants (overconfident and more confident than AI) and no {\it type C} participants (underconfident but more confident than AI) in our experiment. When the AI's accuracy and confidence are lower, there may be more users whose confidence exceeds that of the AI. In such cases, there will be more {\it type A} and {\it C} participants. If {\it type A} participants become the majority, confidence alignment may enhance overall confidence calibration by reducing the overconfidence of many participants.}

We further explored the consequences of changes in human self-confidence calibration. In the paradigm where AI serves as an advisor, that is, in AI-assisted decision making, worsened self-confidence calibration lead people to less frequently adopt AI's correct predictions, resulting in more errors. This is because, for most participants, the reason for the deterioration in calibration was that self-confidence aligned with AI confidence and increased away from their accuracy. Higher self-confidence led these individuals to rely more on their own decisions rather than adopting AI's advice, as described in previous research \cite{chong2022human}. This also explains why, when self-confidence calibration worsened, people were less likely to adopt AI's incorrect predictions, exhibiting a lower over-reliance metric. In the paradigm where AI acts as a peer collaborator, worse self-confidence calibration impaired the decision making mechanism's ability to appropriately adopt human and AI decisions, manifesting experimentally as an increased proportion of decisions adopting the incorrect decision when only the human or AI was correct. This confirmed our concern that poorer human self-confidence calibration resulting from alignment makes it difficult to discern the complementary performances of humans and AI---the mechanism struggles to determine who is more likely to be correct in certain situations.
Then, naturally, we also found that poorer self-confidence is associated with lower accuracy in human-AI decision making.
These findings are consistent with past discussions in research among humans: when team members differ significantly in capabilities, confidence alignment can affect confidence calibration and final decision accuracy \cite{bang2017confidence,Pescetelli2022benefits}.

\subsection{Understanding AI Confidence's Effect on Human's Confidence in Joint Human-AI Decision}
Under the paradigm where AI acts as an advisor, that is, in AI-assisted decision making, we observed that the confidence people have in their joint decisions, which are made after considering AI's advice, aligns much more closely with AI confidence than the alignment between people's initial self-confidence and AI confidence. We believe this is because when people consider AI's advice, they also to some extent adopt the AI's confidence in its own predictions. Our further findings support this conjecture, indicating that when people's final decisions align with AI's recommendations, their confidence in these joint decisions aligns more closely with AI confidence. Past research has explored human self-confidence and confidence in AI's advice in AI-assisted decision making, but has rarely touched upon human confidence in the final joint decision \cite{chong2022human,ma2024you}. We consider that human decision-makers' confidence in joint decisions reflects, to a certain extent, their assessment of their ability to make decisions assisted by AI, which is an important process of their metacognition \cite{yeung2012metacognition}. Our findings offer an interesting perspective on how this metacognitive process is influenced by the confidence of the AI collaborator.

\subsection{Theoretical and Design Implications}
Our study makes significant theoretical contributions. It enriches the literature on the dynamics of human self-confidence in human-AI decision making and on utilizing AI confidence for complementary collaboration \cite{chong2022human,ma2024you,zhang2020effect,lai2023towards}. We demonstrate that in the interaction processes of human-AI decision making, human self-confidence is not independent of AI confidence; rather, it is influenced by and aligns with AI confidence. This addresses the research gap in understanding the effect of AI confidence on human self-confidence. 
Secondly, our study also extends theories of self-confidence interaction among humans, namely confidence alignment, to interactions between humans and AI. This contributes new theoretical insights to modern HCI theory and human metacognition research.
Meanwhile, our study provides empirical evidence for the enduring effect of self-confidence alignment in independent tasks following human-AI interaction, further complementing previous research \cite{Pescetelli2022benefits}.
Moreover, building on prior research on confidence alignment \cite{Pescetelli2022benefits,bang2017confidence}, our study further explores the effect of real-time feedback on confidence alignment, enriching the theories in confidence alignment.

To design effective human-AI decision making systems, our findings suggest that preventing the deterioration of self-confidence calibration caused by the alignment of human self-confidence with AI confidence, along with the associated inappropriate reliance and diminished efficacy of human-AI decision making, is a current necessity. Researchers and developers designing systems and processes for human-AI interactions that include elements of AI uncertainty or confidence must consider the influence of AI confidence on human self-confidence and its potential adverse effects. 
This is crucial because our findings indicate that human self-confidence changes whenever humans observe AI confidence, regardless of the form of collaboration or even the presence of collaboration. 
The mitigating effect of real-time feedback on alignment inspires us: In human-AI interactions, incorporating elements that help humans recognize their capabilities--such as providing feedback about accuracy and displaying social transparency regarding human capabilities in similar tasks \cite{ehsan2021expanding}--could effectively alleviate the issue of human self-confidence calibration caused by confidence alignment.

The alignment of human self-confidence with AI confidence also offers positive implications. 
Similar to scenarios in human group decision making \cite{bang2017confidence}, in cases where human and AI capabilities are comparable, promoting confidence alignment between humans and AI should improve the efficacy of joint human-AI team decisions. 
\rhl{Furthermore, in future systems incorporating elements of AI uncertainty or confidence, it is better to consider users' own ability levels and confidence levels—whether they are overconfident or underconfident—and how their confidence compares to the AI model's. Tailoring the AI's expression of confidence according to different user confidence types (e.g., overconfident users) can achieve better human-AI collaboration. For example, assigning AIs with lower confidence levels to overconfident participants can help calibrate their confidence.}
Additionally, echoing previous confidence alignment research \cite{Pescetelli2022benefits}, in interventions for metacognitive deficits, such as in patients with mental disorders characterized by overconfidence or underconfidence, it might be possible to use an AI agent with complementary confidence levels to interact with the patients and regulate their self-confidence levels and metacognitive abilities.

\subsection{Limitations}
Although our experiment considered various paradigms of human-AI decision making and the presence or absence of real-time feedback, caution is still needed when generalizing to other tasks and experimental groups. 

Firstly, we recognize some limitations to the decision making tasks. We used income prediction that posed low risk and was not complex for non-specialists. It remains unknown whether the same results would be observed in high-risk decision making scenarios or tasks requiring more specialized knowledge, such as medical image recognition \cite{rajpurkar2022ai} or investment decisions \cite{cao2022ai}. 
Our study employed a relative large number of decision-making tasks~\cite{zhang2020effect,rechkemmer2022confidence,ma2023should}, which may introduce a learning effect. Although the presence of a tutorial could mitigate the learning effect, and preliminary analyses have excluded its influence, this remains a noteworthy limitation. 
\rhl{Moreover, we did not control for the degrees of difficulty within each stage. 
Differences in task difficulty may influence confidence changes between consecutive tasks. 
This limitation restricted our ability to observe the effects of specific decision types on human confidence and confidence alignment. For instance, we were unable to examine changes in confidence when the participant's first decision was incorrect, the AI's prediction was correct, and the joint decision was ultimately correct.}

Secondly, we also notice some limitations to the AI model and the method of displaying uncertainty we used. Compared to the currently popular large language models (LLMs) that can communicate uncertainty through language \cite{xiong2023can}, the AI in our experiment could only display a numerical representation of confidence level. Different uncertainty expression forms may make the confidence alignment different. 
\rhl{The fixed accuracy of the AI is also a limitation. The AI's accuracy can influence participants' perceptions of its capabilities. If the AI's accuracy is lower than participants', leading them to believe the AI has poor ability~\cite{rechkemmer2022confidence}, they may disregard the AI's suggestions and confidence levels. Lower AI accuracy also increases errors in decision-making, and the resulting negative feedback can reduce people's self-confidence~\cite{chong2022human}. These could impact the process of confidence alignment.}

\rhl{Meanwhile, we note that a possible confound in the study is that the majority of participants had lower confidence than the AI, so they would increase their confidence for confidence alignment. It is also possible that there are other mechanisms for increasing confidence that may differ between conditions--for example, when the AI serves as a peer collaborator, participants could possibly have higher motivation to increase their own confidence for their answer to be chosen. Our study design would not allow for distinguishing those mechanisms with statistical significance and we encourage future work to explore the interactive effects between human-AI decision paradigms and original differences between human-AI confidence.}

\rhl{Lastly, most participants exhibited relatively high accuracy in this study. As a result, their decisions tended to agree with the AI rather than disagree in our study.
Although the linear correlation between participants' accuracy and confidence alignment was not significant, since we found the influence of disagreement on participants' confidence in joint decision, it is also meaningful to explore whether the phenomenon of self-confidence aligning with AI confidence changes when people disagree more with the AI.
}

\subsection{Future Work}
With the empirical results of this study, future research could undertake extensive theoretical extensions or practical experiments. 
Future research should be providing more direct evidence to explain the phenomenon of confidence alignment. 
Also, in exploring the dynamics of human self-confidence in human-AI decision making, future studies could develop new predictive models that include AI confidence, aiming for a more precise prediction of the dynamics of human self-confidence. 

\rhl{Future studies can further investigate the influence of imperfect AI confidence on confidence alignment. In our study, the AI's mean confidence can reflect its accuracy well. However, when people collaborate with overconfident or underconfident AI, the process of confidence alignment may differ from the results of this study. Because people's perception of AI confidence calibration can affect their trust in the suggestion and confidence reported by AI \cite{li2024overconfident}, which may influence confidence alignment}. 

\rhl{Additionally, some individuals are generally more confident while some are less confident \cite{bang2017confidence}. It is also interesting to investigate how individuals' trait self-confidence~\cite{Pescetelli2022benefits}, as influenced by socioeconomic factors, affects the process of confidence alignment. For instance, will some individuals align their confidence with AI, while others do not? These questions await further exploration.}


With the popularity of LLMs, a common focus in both industry and academia is how to train models so that LLMs align with humans in behaviors, beliefs, and other aspects \cite{achiam2023gpt}. Compared to the studies aligning AI to humans \cite{achiam2023gpt}, exploring the impact of these increasingly powerful AIs on the behavior and perspectives of human users is equally important \cite{kobis2021bad}. 
Our work, along with many others, demonstrates that humans align their cognition, perspectives and behaviors with AI during its use \cite{kobis2021bad,hertz2018under, vollmer2018children, hertz2019mixing, song2024multi}.
Future research could examine the alignment of language expressed uncertainty between humans and AI capable of natural language interaction during collaborative processes, thereby extending the content of this study.
Typically, the alignment during interactions facilitates more effective communication \cite{branigan2006alignment}.
As mentioned in the design implications section of this paper, if harnessed properly, these impacts can yield substantial social goods. 
However, they might also result in negative outcomes, such as eroding moral values \cite{kobis2021bad}. 
Thus, we call on HCI scholars to bring forward more engaging work about the alignment of human cognition and behavior with AI to foster more harmonious human-AI relationships and generate greater societal benefits.

\section{Conclusions}
Human-AI decision making is increasingly being used across various decision-making domains \cite{lai2023towards,yablonsky2021ai}. Inspired by past research \cite{bang2017confidence,Pescetelli2022benefits,fusaroli2012coming}, we point out that in human-AI decision making, AI confidence has a potential influence on human self-confidence, which in turn could further affect the efficacy of human-AI decision making.
In this study, through a randomized behavioral experiment, we found that the phenomenon of human self-confidence aligning with AI confidence is commonly present in human-AI decision making, and it can persist even after AI ceases to be involved. We also revealed that for most users, such alignment could impair their self-confidence calibration. Such poor self-confidence calibration is related closely to inappropriate reliance and low human-AI decision making efficacy.
Our results offer a new perspective for understanding the dynamics of human self-confidence in human-AI decision making, revealing that AI confidence is not merely an indicator of AI performance but can also have more complex and potential influences. 
We hope that future work can build on this study to explore more cognitive influences of AI on users during interaction.

\begin{acks}
This research is supported by the National University of Singapore School of Computing grant (A-8000529-00-00), as well as by the Singapore Ministry of Education Academic Research Fund (A-8002610-00-00). We thank all reviewers’ comments and suggestions to help polish this paper
\end{acks}

\bibliographystyle{ACM-Reference-Format}
\bibliography{sample-base}


\end{document}